\documentclass[prd,twocolumn,nofootinbib]{revtex4}
\usepackage{graphicx} 
\usepackage{epsfig} 
\usepackage{hyperref}

\newcommand{\beq}{\begin{equation}}
\newcommand{\eeq}{\end{equation}}

\begin{document}

\title{Global Structure of the Multiverse and the Measure Problem}

\author{Alexander Vilenkin}
\email{vilenkin@cosmos.phy.tufts.edu}
\affiliation{Institute of Cosmology, Department of Physics and Astronomy, Tufts University, Medford, Massachusetts 02155, USA.}

\begin{abstract}
An unresolved question in inflationary cosmology is the assignment of 
probabilities to different types of events that can occur in the
eternally inflating multiverse. 
We explore the possibility that the resolution of this ``measure
problem" may rely on non-standard dynamics  
in regions of high curvature. In particular, ``big crunch"
singularities that develop in bubbles with  
negative vacuum energy density may lead to bounces, where contraction
is replaced by inflationary expansion.  Similarly, singularities
inside of black holes might be gateways to other inflating vacua.
This would drastically affect the global structure of the 
multiverse. We consider a measure based on a probe geodesic which
undergoes an infinite number 
of passages through crunches. This can be thought of as the world-line
of an eternal ``watcher",  collecting data in an orderly fashion.  
This watcher's measure is independent of initial conditions and does
not suffer from ambiguities associated with the choice of a cut-off
surface.

\end{abstract}

\maketitle

\section{Introduction}

String theory predicts an enormous landscape of vacua with diverse
physical properties.  Extra dimensions allow a multitude of different
compactifications; in addition, fluxes can thread these extra
dimensions and branes can wrap around them in many different ways.
The number of possibilities is combinatorial, and estimates of the
total number of vacua in the landscape go as high as $10^{1000}$
(see \cite{Douglas} for a recent review).  Combining this with
inflationary cosmology leads to the picture of a multiverse, where
bubbles of all possible vacua are constantly being formed in the
course of eternal inflation. (For a review of eternal inflation,
see \cite{Guth07}. 

This picture, however, encounters a serious difficulty, known as the measure problem.  The problem is how to assign probabilities to different events.  In a finite spacetime, this could be done simply by counting, so the relative probability of events $A$ and $B$ would be given by
\beq
\frac{P_A}{P_B}=\frac{N_A}{N_B} ,
\label{PPNN}
\eeq
where $N_A$ and $N_B$ are the corresponding numbers of occurrences.  However, in an eternally inflating universe, any event having a non-vanishing probability will occur an infinite number of times, so the right-hand side of (\ref{PPNN}) is ill defined.

To better explain what the measure problem is and what it is not, consider the sequence of natural numbers,
\beq
1,~2,~3,~4,~5,~6,~7,~8,~...~ , 
\label{123}
\eeq
and let us ask the question: What fraction of these numbers are even?  If we take $N$ numbers in a row, this fraction will be close to $1/2$ for large $N$ and will exactly equal $1/2$ in the limit $N\to\infty$.  However, if we reorder the sequence as
\beq
1,~3,~2,~5,~7,~4,~9,~11,~...~ , 
\label{132}
\eeq
the same procedure would give $1/3$, and it is clear that we can get any answer with a suitable reordering.
In this particular example, the ambiguity can be easily avoided: there is a natural ordering, which is given by (\ref{123}), and we can require that this ordering be used in the limiting procedure.  The answer is then $1/2$, as one could intuitively expect.

One could try to adopt a similar prescription for the probabilities in the multiverse, using the natural ordering of events in time,
\beq
\frac{P_A}{P_B}=\lim_{t\to\infty}\frac{N_A(t)}{N_B(t)} ,
\label{PPNNt}
\eeq
where $N_{A,B}(t)$ are the numbers of times that events $A$ or $B$ occurred prior to time $t$.  The problem is, however, that the result depends on how time is defined (that is, how clocks are synchronized) in different places. There is no unique, or preferred way to do that in GR.  Hence, the relative probability (\ref{PPNNt}) remains ambiguous.  This is the essence of the measure problem.

Even though the measure problem has been mostly discussed in relation to the multiverse, it arises in all versions of inflationary cosmology in which inflation is eternal, that is, in practically all models of inflation that have been considered so far.  The problem was uncovered about 20 years ago by Andrei Linde and his collaborators \cite{lime1,GarciaBellido} and still remains unresolved.  This suggests that some important element may be missing in our understanding of eternal inflation.  Here, I am going to argue that the key may be in the physics of super-dense regions that develop as a result of gravitational collapse in negative energy vacua.
This is based on my recent work with Jaume Garriga \cite{GV12}, where we suggest that instead of a big crunch singularity, the collapse may be followed by a bounce with a subsequent inflationary expansion.   We shall see that such non-singular bounces open a new approach to the measure problem.

The paper is organized as follows.  In the next section I briefly
review some of the measure prescriptions that have been suggested so
far.  Nonsingular bounces, the resulting global structure of
spacetime are discussed and the new measure proposal is introduced in Section
3.  Section 4 presents a formalism for calculating
probabilities in the new measure.  In Sections 3 and 4 I assume that
the landscape does not include any stable Minkowski vacua; the effects
of such vacua and of black hole nucleation are is discussed in Section 5.  Finally, the conclusions are
briefly summarized in Section 6.

\section{Global and local measures}

The measure problem has been extensively discussed in the literature.
Here, I omit most of the references, keeping only the most relevant
ones and referring the reader to the recent review \cite{Freivogel}
for more details.  For definiteness, I will confine the discussion to
the case of a multiverse, where transitions between different vacua
occur through bubble nucleation.  Bubbles of daughter vacua nucleate
in the inflating background of the parent vacuum and expand, rapidly
approaching the speed of light. 
The bubbles can be divided into three classes: (i) positive-energy or de Sitter (dS), (ii) negative-energy or Anti-de Sitter (AdS), and (iii) zero-energy or Minkowski bubbles.  Inflation continues in dS bubbles, but eventually comes to a stop in AdS and Minkowski bubbles.  Hence, the latter two types of bubbles are called "terminal bubbles".\footnote{I assume for simplicity that the Minkowski vacua are stable, which is certainly the case if they are supersymmetric.}

The simplest way to regulate the infinite numbers of events is to introduce a global time cutoff.  The starting point is a congruence of timelike geodesics orthogonal to a smooth spacelike hypersurface $\Sigma_0$.  (This hypersurface does not have to be infinite.)  A time coordinate $t$ is defined along the geodesics, with $t$ growing towards the future and $t=0$ at the point of intersection with $\Sigma_0$.  The hypersurfaces $\Sigma(t)$ of constant $t$ (which are not necessarily spacelike) can then be used as cutoff surfaces for the measure.  Specifically, we can define $N_A(t)$ and $N_B(t)$ in Eq.~(\ref{PPNNt}) as the corresponding numbers of events that occurred in the comoving spacetime region between the hypersurfaces $\Sigma_0$ and $\Sigma(t)$.   The relative probabilities of $A$ and $B$ can then be found by taking the limit (\ref{PPNNt}).

The choice of the time parameter $t$ is largely arbitrary.  The most popular choices are the proper time $\tau$ and the scale factor $a$ along the geodesics.  The probability distributions resulting from this procedure are independent of the initial surface $\Sigma_0$, but are very sensitive to the the choice of $t$.  For example, with the proper time cutoff a higher expansion rate of inflation is exponentially favored, while the scale factor cutoff gives no preference to fast expansion.

Apart from this lack of uniqueness, there are also problems of a more technical character.  Geodesic congruences tend to develop caustics; then the time variable $t$ becomes multi-valued.  Moreover, some of the time variables that are commonly used are not monotonic.  (For example, the scale factor decreases along 
the geodesics in contracting parts of the universe.)  One needs to introduce additional rules to handle these cases.

Another class of measures, introduced by Raphael Bousso and his collaborators, includes the so-called local measures, which
sample a spacetime region  
in the vicinity of a given timelike geodesic.  Here again, there are a
number of possible choices for the sampling region.  It could, for example, be the
past light cone of the geodesic (the causal patch measure \cite{Bousso1}),
 or the region within a fixed physical distance of the geodesic (the `fat geodesic'
measure \cite{Bousso2}).  

A related proposal, closer in spirit to the one we shall explore here, is that instead of counting
observations made by all observers within a spacetime region 
defined by a geometric cutoff, we include only observations made by a
single `observer' specified by a timelike geodesic.  A simple version of such 
a measure was introduced in \cite{GV98} and later
discussed in \cite{VV,Vanchurin}.  Most recently, the single observer picture was
discussed   
by Nomura \cite{Nomura}, who motivated it from quantum mechanical considerations. 

The basic problem of all local measures, including the single observer measure, 
was pointed out already in \cite{GV98}.
A typical geodesic, starting in some inflating de Sitter (dS) vacuum, 
will traverse a number of dS bubbles and
will eventually enter a terminal bubble -- either an AdS bubble terminating at a big crunch, or a bubble of
supersymmetric stable Minkowski vacuum.  All geodesics, except for a set of measure zero,
will then visit a finite number  of
bubbles, so the resulting probability distribution will depend on what
geodesic we choose.  Hence, one needs to consider
an ensemble of geodesics with different initial conditions.  Without
specifying such an ensemble, these measures remain essentially
undefined. 

Much of the recent work on the measure problem has been aimed at 
exploring phenomenological aspects of different measure proposals, making sure they are not riddled
with internal inconsistencies or obvious conflict with the data (see \cite{Freivogel} for a review).  
Some of the measure candidates have already been ruled out in this way, but it seems unlikely that this sort of phenomenological analysis will yield a unique measure prescription.

\section{Nonsingular bounces and a measure proposal}

Thus, despite a considerable effort, the measure problem remains
unresolved.  This motivated Jaume Garriga and me to explore a different approach, which assumes that the global structure of the multiverse significantly differs from the standard picture \cite{GV12}.  Specifically, we
conjecture that spacetime singularities will eventually be resolved in the fundamental
theory of Nature, so that the big crunches that occur in AdS bubbles 
will turn out to be nonsingular.  The
standard description of AdS regions would then be applicable at the
initial stages of the collapse, but when the density and/or curvature
get sufficiently high, the dynamics would change, resulting in a
bounce.  Scenarios of this sort have been discussed in the 1980's in
the context of the so-called maximum curvature hypothesis
\cite{FMM1,FMM2}, and more recently in the context of pre-big-bang scenario \cite{Veneziano},  ekpyrotic and cyclic models \cite{ekpyrotic,cyclic}, loop quantum cosmology \cite{Ashtekar,Bojowald} and holographic ideas \cite{Brustein}.\footnote{For a recent discussion of nonsingular bounces in the context of eternal inflation, see Refs.~\cite{Piao1,Piao2,JL,J-L}.}  The only terminal vacua
in this picture are stable Minkowski vacua. 

I will argue that this global structure allows for an improvement in the
definition of the measure.  To explain the idea, 
let us first assume that the landscape does not include stable
Minkowski vacua.  (A more general case will be discussed in Section 5.) The corresponding spacetime structure is illustrated in Fig.~1.  The would be singularities, which are now replaced by bounces, are indicated by zigzag lines.  Because of the high energy densities reached near the bounce, the crunch regions are likely to be excited above the
energy barriers between different vacua, so transitions to other vacua
are likely to occur.  We shall make no assumptions about the dynamics
of the bounce and simply characterize AdS vacua by the
transition probabilities to new (dS or AdS) vacua after the crunch.
Different parts of the same crunch region can, of course, transit to
different vacua.  

\begin{figure}
  \includegraphics[height=.25\textheight,clip=true]{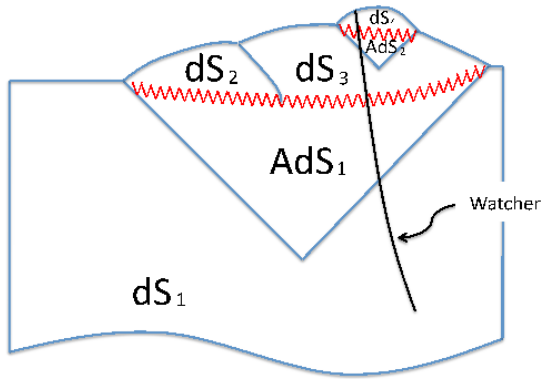}
  \caption{Causal diagram of a multiverse with both positive (dS) and negative (AdS) energy density regions. 
The worldline of the watcher goes through an infinite sequence of AdS crunches.}
\end{figure}


All future-directed timelike geodesics in this spacetime pass through a succession of dS and AdS vacua,
extending all the way to future infinity.  Given a finite number of
vacua in the landscape, a generic geodesic will pass through each
vacuum an infinite number of times.\footnote{Here we assume that the vacuum landscape is irreducible, i.e. any vacuum can be reached by a sequence of transitions from any other vacuum.} 
The number of distinguishable events that can be detected by an
"eternal observer" following such a geodesic is also finite, and thus
any event that has a nonzero probability will be observed an infinite
number of times \cite{manyworlds}.  Our measure proposal is that {\it
the probabilities of different events are proportional to the
frequencies at which these events are encountered by an eternal
observer.} 
To avoid confusion between the eternal observer and physical observers in the multiverse, we shall refer to the eternal observer as "the watcher".

We now have to spell out what exactly is meant by events "encountered" by a geodesic.  "Events" in General Relativity are often represented by points in spacetime.  However, macroscopic 
events that are of interest to us are extended in both space and time.
Hence, we shall assume that each type of event $A$ can be characterized by
a finite domain $D_A$, defined as the minimal spacetime region that is
necessary to specify that event, and by a cross-section $\sigma_A$
that it presents for the watcher's geodesic.  We could then count only events
whose domain is traversed by the geodesic. 

As it stands, this prescription is not quite satisfactory, since it
gives preference to events with a large cross-section.  For example, a
measurement that uses bulky equipment or takes a large amount of time
will be assigned a higher probability.  In order to correct for this
effect, we shall introduce the corrected number of encounters $N_A$, 
\beq
N_A = \frac{\sigma_0}{\sigma_A} \nu_A, \label{corrected}
\eeq
where $\nu_A$ is the number of passages through domains of type $A$
and $\sigma_0$ is an arbitrary constant with the dimension of volume.
The relative probability of 
events $A$ and $B$ is then given by Eq.~(\ref{PPNNt}) with 
$N_A(t)$ and $N_B(t)$ being the corresponding numbers of encounters
up to time $t$ along the geodesic.  Note that the situation here is
similar to that with the natural numbers: the events are counted in
the order as they occur.  It does not matter which time variable $t$
is used, as long as it is monotonic along the geodesic.  Also, it does
not matter which geodesic we choose.  The same probability
distribution will be obtained for all geodesics, except a set of
measure zero.   

An attractive property of the watcher measure is that it does not require an arbitrary choice of the cutoff surface or an arbitrary specification of an ensemble of geodesics.  The phenomenology of this measure has not yet been studied in detail.  A preliminary analysis in Ref.~\cite{GV12} indicates that it is qualitatively similar to the scale factor cutoff measure.  In particular, it gives a correct prediction for the cosmological constant and appears to be free of paradoxes afflicting some other measures.\footnote{The only exception appears to be the Guth-Vanchurin paradox, which is concerned with measurements extending in time over periods comparable to or longer than the Hubble time.  Whether or not this paradox is a serious problem is a matter of some debate.  Guth and Vanchurin
\cite{guthvanchurin} argued that it is just a peculiar feature of probabilities in an eternally inflating universe and does not cause any phenomenological problems, while Olum \cite{ken} argued that this feature violates some requirements that any definition of probability should satisfy.}

\section{Frequency of visits and the arrow of time}

The history of a watcher can be described using a discrete time
variable, $n=1,2,3,...$, which is incremented by one whenever the
watcher jumps to a different vacuum.   The probability $X_i(n)$ to
visit vacuum $i$ at 'time' $n$ should then satisfy the rate
equation \cite{GSVW,VV} 
\beq
X_i(n+1)=\sum_j T_{ij}X_j(n),
\label{rateeq}
\eeq
where $T_{ij}$ is the probability of visiting vacuum $i$ after vacuum
$j$ and the summation is over all vacua in the landscape. 

It is shown in \cite{GV12} that in the limit of large $n$, $X_j(n)$
approaches a unique stationary distribution $X_j$, which satisfies 
\beq
X_i=\sum_j T_{ij}X_j.
\label{eigeneq}
\eeq
The quantities $X_j$ are the frequencies of visits to the corresponding vacua.  Eq.~(\ref{eigeneq}) says that the distribution $X_j$ can be found by solving an eigenvalue problem: it is an eigenvector of the transition matrix $T_{ij}$ with a unit eigenvalue.

If $j$ is a dS vacuum, then the transition probability $T_{ij}$ can be
expressed as 
\beq
T_{ij} = \Gamma_{ij}/\Gamma_j, ~~~~~ \Gamma_j=\sum_i \Gamma_{ij},
\eeq
where 
\beq
\Gamma_{ij}\propto V_j e^{-I_{ij}} e^{-S_j}
\eeq
is the transition rate per unit proper time from vacuum $j$ to vacuum
$i$, $I_{ij}$ is the Coleman-DeLuccia \cite{CdL} instanton action for the
tunneling from $j$ to $i$, 
$V_j=(4\pi/3)H_j^{-3}$ is the horizon volume, $S_j=\pi/H_j^2$ is the Gibbons-Hawking entropy, and $H_j$ is the Hubble expansion rate in vacuum $j$.  
If both vacua $i$ and $j$ are dS, the transition rates satisfy
\beq
T_{ij}/T_{ji}= (V_j/V_i) \exp(S_i-S_j),
\label{detailedbalance}
\eeq
where we have used the fact \cite{LeeWeinberg} that the transitions $i\to j$ and $j\to i$ are both described by the same instanton.
Noting that $\exp(S_j)$ is the number of quantum states in (a horizon region of) vacuum $j$, we recognize that, apart from the pre-exponential factor, Eq.~(\ref{detailedbalance}) is the detailed balance condition.  

In a purely dS landscape, this condition allows an exact solution of the eigenvalue problem (\ref{eigeneq}).  Up to a normalization factor, the solution is \cite{VV}
\beq
X_j\propto \Gamma_j V_j^{-1} \exp(S_j).
\eeq
The average time spent by the watcher in vacuum $j$ is $\tau_j=\Gamma_j^{-1}$, and thus the fraction of time spent in that vacuum is
\beq
f_j \propto \tau_j X_j \propto V_j^{-1} \exp(S_j).
\eeq
Apart from the factor $V_j^{-1}$, this is the microcanonical
distribution: the fraction of time spent in vacuum $j$ is proportional
to the number of quantum states in that vacuum.  This suggests that in
a purely dS multiverse the watcher will observe a state of thermal
equilibrium, which means in particular that the most probable way of
getting to the state that we now observe is by a 'thermal' (quantum)
fluctuation from an empty, low-energy dS universe. The probability of
tunneling up to a high-energy vacuum with subsequent inflation and
standard big bang evolution is negligibly small by comparison.  This
picture, which was first discussed by Dyson, Kleban and
Susskind \cite{Dyson}, indicates that we do not live in a pure dS
multiverse. 

 For an AdS vacuum $j$, the transition matrix elements $T_{ij}$ are determined by the unknown dynamics of AdS bounces.  However, there seems to be no reason why these matrix elements should satisfy the detailed balance condition (\ref{detailedbalance}).   Hence, we expect that AdS bounces induce departures from thermal equilibrium and thus introduce an arrow of time.

\section{Minkowski vacua and black holes}

Suppose now that the landscape includes some stable supersymmetric
Minkowski vacua 
(we shall call them $M$-vacua), as suggested by string theory.  Then
all timelike geodesics, except a set of measure zero, will end up in
these terminal vacua, with their endpoints at Minkowski timelike
infinity (see Fig.~2).  We shall refer to them as $M$-geodesics. The
remaining, measure-zero geodesics which are constantly recycling
between different vacua will be called $R$-geodesics.  The set of
vacua visited by any $M$-geodesic is finite and depends on where the
geodesic started.  Hence, $M$-geodesics are not useful for defining
probabilities (unless we re-introduce an ensemble with some
distribution of initial data). 

\begin{figure}
  \includegraphics[height=.25\textheight, clip=true]{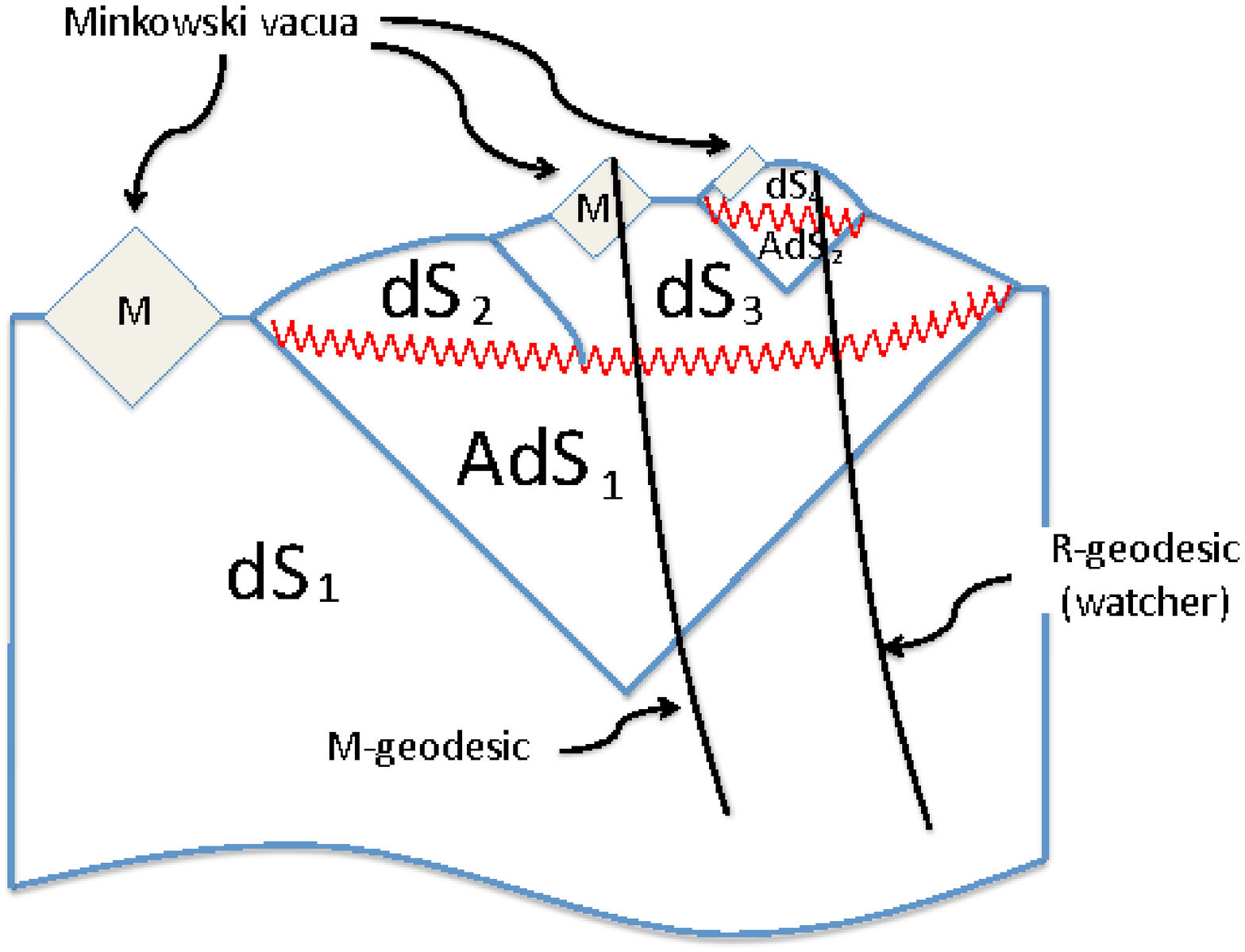}
  \caption{Causal diagram for a multiverse with AdS bounces and terminal Minkowski vacua.}
\end{figure}


The strategy that suggests itself in this case is to determine probabilities by following one of $R$-geodesics.\footnote{Note that this approach is in some sense opposite to that adopted in the census taker measure \cite{census}, which is focused entirely on $M$-geodesics.}
It does not matter which geodesic we choose.  
This prescription can be justified by observing \cite{SusskindBook} that supersymmetric $M$-vacua cannot support nontrivial chemistry, and thus no measurements can take place in such vacua.  

Even though $R$-geodesics do not end at Minkowski timelike infinity, they can visit Minkowski ($M$) bubbles.  To see how this can happen, we note that early evolution of $M$-bubble interiors may include periods of inflation and matter domination, during which supersymmetry will be broken.   During such periods, bubbles of other vacua can form and expand.  The spacetime structure of such bubbles is somewhat unusual.  Suppose, for example, that there is an early period of inflation in the $M$-bubble and that a daughter dS bubble has nucleated during this period.  Initially, the daughter bubble will expand, just as it would in a parent dS vacuum.  It will continue to expand even after the background energy density drops below that in the daughter bubble -- as long as the bubble radius remains larger than the local horizon.  But eventually this condition is violated, so
the daughter bubble begins to contract and collapses to a black hole.  
In the meantime its interior continues to expand and to form its own daughter bubbles.  After the black hole eventually evaporates, this interior becomes a separate inflating multiverse.  

We thus see that the multiverse generally has a rather complicated spacetime structure, and includes a multitude of spatially disconnected regions.\footnote{This spacetime structure is similar to that discussed by Sato, Kodama, Sasaki and Maeda \cite{Sasaki} in a somewhat different context.}  The watcher's geodesic can transit from one such region to another by entering an $M$-bubble and exiting through one of its daughter bubbles. Each $R$-geodesic will generally visit an infinite number of disconnected regions.

If AdS singularities are resolved in the future fundamental theory, the same is likely to apply to black hole singularities.  
Black holes can spontaneously nucleate in de Sitter space at a certain rate per unit spacetime volume (see, e.g., \cite{BoussoHawkingBH} and references therein), and thus the watcher's geodesic has some probability to encounter a black hole per unit proper time.  
As it enters a black hole, the geodesic hits the high-curvature region
replacing the singularity and transits to another dS or AdS vacuum.

For black holes smaller than the horizon, $M\ll H^{-1}$, the
nucleation rate per unit spacetime volume is given by 
\beq
\Gamma_{BH}(M) \propto e^{-M/T_{GH}} ,
\label{GammaBH}
\eeq
where $M$ is the black hole mass and $T_{GH} = H/2\pi$ is the
Gibbons-Hawking temperature of de Sitter space.  This grows
exponentially as $M$ is decreased, indicating that the watcher is most
likely to be captured by a minimal black hole, having size determined
by the UV cutoff of the theory.  The probabilities will then have a
strong dependence on UV physics.  Note,however, that in some
approaches to quantum gravity (e.g., \cite{Nomura}) the fundamental
entity is the wave 
function of a region encompassed by an apparent horizon.  The watcher
is then effectively defined by the apparent horizon surface, rather
than by a geodesic.  Nucleation of small black holes has little effect
on the apparent horizon, so one can expect that the measure will not
be UV-sensitive.

Finally, I would like to comment on an alternative possibility for the
measure.\footnote{I am grateful to Jaume Garriga for suggesting this
possibility and for subsequent clarifying discussions.}  Suppose the
crunches in AdS bubbles and in black hole 
interiors are singular, so that geodesics cannot be continued beyond
the crunches.  Then, we could still define a measure using
$R$-geodesics, which extend to future infinity 
without ever encountering $M$-bubbles, AdS bubbles, or black holes.
It appears, however, that with this prescription the watcher's
geodesic will not adequately sample habitable regions at early stages in $M$-bubbles and in AdS bubbles with small (negative) values of the cosmological constant.\footnote{The watcher's geodesic can visit $M$ bubbles and exit through dS bubbles that form inside, as discussed earlier in this section.    However, as the background matter density in the $M$-bubble goes down, the rate of dS bubble nucleation decreases exponentially.  The probabilities of events in low energy density regions will be suppressed correspondingly.  A similar suppression will occur for events in AdS bubbles.}  Thus, at this stage, the picture of AdS and black hole bounces appears to be more promising.


\section{Summary}

We have reconsidered the measure problem of inflationary cosmology, by introducing the non-standard assumption that
spacetime singularities are resolved in the fundamental theory, in such a way that all timelike geodesics can be extended
indefinitely into the future. A measure can then be defined using a single future-eternal timelike geodesic. This geodesic 
can be thought of as the worldline of a ``watcher", sampling different types of events as they are intersected in the course of time.  This measure is independent of initial conditions and does not involve an arbitrary choice of a sampling cutoff region.  
Phenomenologically, this measure is similar to the scale factor cutoff measure. 
Since the latter does not suffer from any obvious phenomenological problems, we expect the watcher's measure to do just as well.

\section*{Acknowledgments}

I am grateful to Jaume Garriga for collaboration when most of the work reported here was done and for very useful discussions during the preparation of this paper.  I also got useful feedback on these ideas from the participants of the Long-term Workshop YITP-T-12-03 on "Gravity and Cosmology 2012" at the Yukawa Institute for Theoretical Physics at Kyoto University.  This work was supported by grant from the Templeton Foundation and by PHY-0855447 from the National Science Foundation. 


\end{document}